\documentclass[11pt]{article}
\usepackage{tipa}
\usepackage[all,import]{xy}
\usepackage{graphicx,color}
\usepackage{amsmath}
\usepackage{amsbsy}
\usepackage{amssymb}
\usepackage{cases}
\usepackage{enumerate}
\usepackage{bm}
\usepackage[colorlinks,linkcolor=black,anchorcolor= black,citecolor= black,urlcolor=black]{hyperref}
\usepackage[round]{natbib}

\usepackage{amsthm}

\usepackage{times}
\usepackage{mathptmx}

\makeatletter
\newcommand*{\rom}[1]{\expandafter\@slowromancap\romannumeral #1@}
\makeatother

\begin{document}

\theoremstyle{definition}
\newtheorem{assumption}{Assumption}
\newtheorem{theorem}{Theorem}
\newtheorem{lemma}{Lemma}
\newtheorem{example}{Example}
\newtheorem{definition}{Definition}
\newtheorem{corollary}{Corollary}

\def\letas{\mathrel{\mathop{=}\limits^{\triangle}}}
\def\ind{\begin{picture}(9,8)
         \put(0,0){\line(1,0){9}}
         \put(3,0){\line(0,1){8}}
         \put(6,0){\line(0,1){8}}
         \end{picture}
        }
\def\nind{\begin{picture}(9,8)
         \put(0,0){\line(1,0){9}}
         \put(3,0){\line(0,1){8}}
         \put(6,0){\line(0,1){8}}
         \put(1,0){{\it /}}
         \end{picture}
    }

\def\AVar{\text{AsyVar}}
\def\Var{\text{Var}}
\def\Cov{\text{Cov}}
\def\sumn{\sum_{i=1}^n}
\def\summ{\sum_{j=1}^m}
\def\convergeas{\stackrel{a.s.}{\longrightarrow}}
\def\converged{\stackrel{d}{\longrightarrow}}
\def\iidsim{\stackrel{i.i.d.}{\sim}}
\def\indsim{\stackrel{ind}{\sim}}
\def\asim{\stackrel{a}{\sim}}
\def\d{\text{d}}

\setlength{\baselineskip}{2\baselineskip}

\title{\bf Semiparametric Inference of the Complier Average Causal Effect with
Nonignorable Missing Outcomes}
\author{Hua Chen,
\footnote{Institute of Applied Physics and Computational Mathematics, Beijing, 100088, China}\quad
Peng Ding,
\footnote{Department of Statistics, Harvard University, One Oxford Street, Cambridge, MA 02138, USA.} 
\footnote{Corresponding author's Email:
\url{pengding@fas.harvard.edu}}
\quad 
Zhi Geng,
\footnote{School of Mathematical Sciences, Peking University, Beijing, 100871, China}\quad 
and Xiaohua Zhou
\footnote{Department of Biostatistics, University of Washington, and Biostatistics Unit, HSR\&D Center of Excellence, 
VA Puget Sound Health Care System, Seattle, Washington 98101, USA}
}
\date{}
\maketitle

\begin{abstract}
Noncompliance and missing data often occur in randomized trials, which complicate the inference of causal effects. When both noncompliance and missing data are present, previous papers proposed moment and maximum likelihood estimators for binary and normally distributed continuous outcomes under the latent ignorable missing data mechanism. However, the latent ignorable missing data mechanism may be violated in practice, because the missing data mechanism may depend directly on the missing outcome itself. Under noncompliance and an outcome-dependent nonignorable missing data mechanism, previous studies showed the identifiability of complier average causal effect for discrete outcomes. In this paper, we study the semiparametric identifiability and estimation of complier average causal effect in randomized clinical trials with both all-or-none noncompliance and the outcome-dependent nonignorable missing continuous outcomes, and propose a two-step maximum likelihood estimator in order to eliminate the infinite dimensional nuisance parameter. Our method does not need to specify a parametric form for the missing data mechanism. We also evaluate the finite sample property of our method via extensive simulation studies and sensitivity analysis, with an application to a double-blinded psychiatric clinical trial.

\bigskip 
\noindent {\bfseries Key Words:} Causal inference;
Instrumental variable; Missing not at random; Noncompliance; Outcome-dependent missing; Principal
stratification.
\end{abstract}

\section{Introduction}
\label{sec1}

Randomization is an effective way to study the average causal effects ($ACE$s) of new drugs or training programs.
However, randomized trials are often plagued with noncompliance and missing data, which may make statistical inference difficult and biased. The noncompliance problem happens when some subjects fail to comply with their assigned treatments, and the missing data problem happens when investigators fail to collect information for some subjects.
Ignoring noncompliance and missing data problems may lead to biased estimators of the $ACE$s.

The noncompliance problem has attracted a lot of attention in
the literature. \citet{Efron1991} studied the noncompliance
problem before the principal stratification framework (\citealp{Frangakis2002}) was proposed. In the presence of noncompliance, \citet{Balke1997} proposed large sample bounds of the $ACE$s for
binary outcomes using the linear programming method. \citet{Angrist1996} discussed the identifiability of the causal effect using the
instrumental variable method. \citet{Imbens1997} proposed a
Bayesian method to estimate the complier average causal effect
($CACE$). When some outcomes are missing, the identifiability and
estimation of $CACE$ are more complicated, and different types of
missing data mechanisms have sizable impacts on the identifiability and
estimation of $CACE$. \citet{Frangakis1999} established the
identifiability and proposed a moment estimator of $CACE$ under the
latent ignorable (LI) missing data mechanism. Under the LI missing
data mechanism, \citet{Zhou2006} and  \citet{Malley2005}
proposed Expectation-Maximization (EM) algorithms (\citealp{Dempster1977}) to find the maximum likelihood estimators (MLEs) of $CACE$ for
binary and normally distributed outcomes, respectively.
 \citet{Barnard2003} proposed a Bayesian approach to
estimate $CACE$ with bivariate outcomes and covariate adjustment. \citet{Taylor2011} proposed a multiple imputation method to estimate $CACE$ for clustered encouragement design studies.

However, the LI assumption may be implausible in some clinical
studies when the missing data mechanism may depend on the missing
outcome. \citet{Chen2009} and  \citet{Imai2009} discussed the
identifiability of $CACE$ for discrete outcomes under the outcome-dependent nonignorable (ODN) missing data mechanism.
To the best of our knowledge,
there are no published
papers in the literature studying the identifiability of $CACE$
for continuous outcomes under the ODN assumption. In this paper, we
show that $CACE$ is semiparametrically identifiable under some
regular conditions, and propose estimation methods for $CACE$ with
continuous outcomes under the ODN assumption. For our semiparametric
method, we need only assume that the distribution of the outcomes
belongs to the exponential family without specifying a parametric form for the missing
data mechanism.

This paper proceeds as follows. In Section \ref{sec::notation}, we discuss the
notation and assumptions used in this paper and define the
parameter of interest. In Section
\ref{sec::semipara}, we show the semiparametric identifiability and
propose a two-step maximum likelihood estimator (TSMLE). In Section
\ref{sec::simulation}, we use several simulation
studies to illustrate the finite sample properties of our proposed
estimators and consider sensitivity analysis to assess the
robustness of our estimation strategy. In Section \ref{sec::data},
we analyze a double-blinded randomized clinical trial using the
methods proposed in this paper.
We conclude with a discussion and provide all proofs 
in the Appendices.

\section{Notation and Assumptions}
\label{sec::notation}

We consider a randomized trial with a continuous outcome. For the $i$-th subject, let $Z_i$ denote the randomized
treatment assignment ($1$ for treatment and $0$ for control). Let
$D_i$ denote the treatment received ($1$ for treatment and $0$ for control). When $Z_i \not= D_i$, there
exists noncompliance. Let $Y_i$ denote the outcome variable.  Let $R_i$ denote the missing data
indicator of $Y_i$, i.e., $R_i=1$ if $Y_i$ is observed and $R_i=0$
if $Y_i$ is missing. First, we need to make the following fundamental
assumption.

\begin{description}
\item[Assumption 1 (Stable unit treatment value
assumption, SUTVA):]
There is no interference between units, which means that the
potential outcomes of one individual do not depend on the treatment
status of other individuals (\citealp{Rubin1980}), and there is only
one version of potential outcome of a certain treatment
(\citealp{Rubin1986}).
\end{description}
Except in the dependent case for infectious diseases (\citealp{Hudgens2008}), the SUTVA assumption is reasonable in many cases.
Under the SUTVA assumption, we define $D_i(z), Y_i(z)$ and $R_i(z)$ as the
potential treatment received, the potential outcome measured, and the potential missing data indicator for subject $i$ if he/she were assigned to treatment $z$.
These variables are potential outcomes because only one of the pairs $\{ D_i(1), Y_i(1), R_i(1) \}$ and
$\{ D_i(0), Y_i(0), R_i(0) \}$ can be observed. Since $Z_i$ is the observed treatment assignment for subject $i$, $D_i = D_i(Z_i), Y_i = Y_i(Z_i)$, and $R_i = R_i(Z_i)$ are the observed treatment received, the observed outcome, and the observed missing data indicator.

Under the principal stratification framework (\citealp{Angrist1996}; \citealp{Frangakis2002}), we let $U_i$ be the
compliance status of subject $i$, defined as follows:
\begin{eqnarray}
U_i=\left \{ \begin{array} {ll}
a, & \text{if} \hspace{2mm} D_i(1)=1 \hspace{2mm} \text{and} \hspace{2mm} D_i(0)=1; \\
c, & \text{if} \hspace{2mm} D_i(1)=1 \hspace{2mm} \text{and} \hspace{2mm} D_i(0)=0; \\
d, & \text{if} \hspace{2mm} D_i(1)=0 \hspace{2mm} \text{and} \hspace{2mm} D_i(0)=1; \\
n, & \text{if} \hspace{2mm} D_i(1)=0 \hspace{2mm} \text{and} \hspace{2mm} D_i(0)=0;
\end{array}
\right. \nonumber
\end{eqnarray}
where $a,c,d$ and $n$ represent ``always-taker'', ``complier'',
``defier'' and ``never-taker'', respectively. Here $U_i$ is an
unobserved variable, because we can observe only $D_i(1)$ or
$D_i(0)$ for subject $i$, but not both. The $CACE$
of $Z$ to $Y$ is the parameter of interest, defined as
$$
CACE(Z\rightarrow Y) = E\{ Y(1) - Y(0)\mid  U=c\}.
$$

$CACE$ is a subgroup causal effect for the compliers, with incompletely observed compliance status.
Next, we give some sufficient conditions about the latent variables to make $CACE(Z\rightarrow Y)$ identifiable, in the presence of noncompliance and nonignorable missing outcomes.

\begin{description}
\item[Assumption 2 (Randomization):]  
The treatment assignment $Z$ is completely randomized.
\end{description}

Randomization means that $Z$ is independent of $\{D(1), D(0),
Y(1), Y(0), R(1), R(0)  \}$, and we define $\xi = P\{ Z=1\mid  D(1), D(0),
Y(1), Y(0), R(1), R(0) \} = P(Z=1)$. Under the randomization assumption, $CACE(Z\rightarrow Y)$ can be expressed as
$$
CACE(Z\rightarrow Y) = E(Y\mid Z=1, U=c) - E(Y\mid Z=0, U=c).
$$

\begin{description}
\item[Assumption 3 (Monotonicity):]  
$D_i(1) \geq D_i(0)$ for each subject $i$.
\end{description}

The monotonicity of $D_i(z)$ implies that there are no defiers.
Define $\omega_u = P(U=u)$ for $u=a,c,d,n$, and the monotonicity assumption implies $\omega_d=0$. Assumption 3
is plausible when the treatment assignment has a nonnegative effect on the treatment received
for each subject, and it holds directly when the treatment is not
available to subjects in the control arm, meaning $D_i(0)=0$
for all subjects. The monotonicity assumption implies a positive $ACE$
of $Z$ on $D$. However, under general circumstances, Assumption
3 is not fully testable, since only one of $D_i(1)$
and $D_i(0)$  can be observed.

\begin{description}
\item[Assumption 4:]
$ACE(Z\rightarrow D)\not=0.$
\end{description}
By randomization, we have that $ACE(Z\rightarrow D)= P(D=1\mid Z=1)
- P(D=1\mid Z=0)\not=0$ under Assumption 4, and therefore $Z$ is
correlated with $D$. Without Assumption 4, we have
$P(D=1\mid Z=1) = P(D=1\mid Z=0)$, which implies that $\omega_c = 0$
under Assumption 3. Since we are interested in the
identifiability of $CACE(Z\rightarrow Y)$, Assumption
4 is necessary.

\begin{description}
\item[Assumption 5 (Compound exclusion restrictions):]
For never-takers and always-takers, we assume $P\{ Y(1), R(1) \mid U=n \} =
P\{ Y(0), R(0) \mid U=n \}$, and $P\{ Y(1), R(1) \mid U=a \} =
P\{ Y(0), R(0) \mid U=a  \} $.
\end{description}

The traditional exclusion restriction assumes $P\{ Y(1)  \mid U=u \} =
P\{ Y(0)  \mid U=u \}$ for $u=a$ and $n$. \citet{Frangakis1999} extended it to the compound exclusion restrictions, and imposed similar assumption on the joint vector of the outcome and the missing data indicator. Assumption 5 is reasonable in a double-blinded clinical trial, because the patients do not know the treatment assigned to them and thus $Z$ has no ``direct effect'' on the outcome and the missing data indicator.
However, when the missing data indicator depends on the treatment assigned, the compound exclusion restrictions may be violated.
When $Z$ is randomized, Assumption 5 is equivalent to $ P(Y, R \mid Z=1,U=n  ) =
P(Y, R \mid Z=0, U=n   ) $ and $ P(Y, R \mid Z=1,U=a ) = P(Y, R \mid
Z=0, U=a   ). $

\begin{description}
\item[Assumption 6 (Outcome-dependent nonignorable missing data):] 
For all $y;$ $z=0,1$; $d=0,1;$ and $u\in \{a,c,n \}$, we assume
\begin{eqnarray}
P\{ R(z)=1 \mid  Y(z) = y, D(z)=d, U=u \} &=& P\{ R(z)=1 \mid  Y(z) =
y\} \label{eq:indeRY1} \\
P\{ R(1) =1 \mid Y(1) = y \} &= & P\{ R(0)=1 \mid Y(0)= y \} .
\label{eq:indeRY2}
\end{eqnarray}
\end{description}
When $Z$ is randomized, the equation \eqref{eq:indeRY1} becomes $ P(R=1 \mid Y=y
,D=d, U=u, Z=z) = P(R=1 \mid Y=y, Z=z),$ and \eqref{eq:indeRY2}
becomes $P(R=1\mid Y=y, Z=1) = P(R=1\mid Y=y, Z=0)$. Define
$\rho(y)=P(R=1\mid Y=y)$. Therefore Assumption 2 and
Assumption 6 imply that $\rho(y)=P(R=1 \mid Y=y, D=d,
U=u, Z=z)$. Hence $R$ depends on $Y$, but
is independent of $(Z,D,U)$ given $Y$.

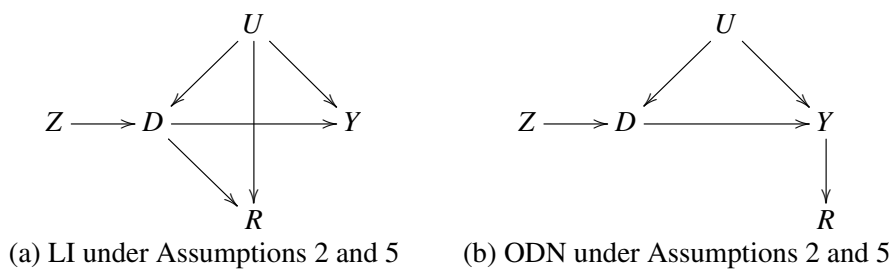
\begin{figure}[ht]
\centering
\begin{tabular}{ccc}
$$
\begin{xy}
\xymatrix{
 & & U \ar[dl] \ar[dr] \ar[dd]\\
Z \ar[r] & D\ar[dr] \ar[rr] & & Y\\
& &R }
\end{xy}
$$
&&
$$
\begin{xy}
\xymatrix{
 & & U \ar[dl] \ar[dr] \\
Z \ar[r] & D \ar[rr] & & Y  \ar[d] \\
&& &R
}
\end{xy}
$$ \\
(a) LI under Assumptions 2 and 5& & (b) ODN under Assumptions 2 and 5
\end{tabular}
\caption{Graphical models for different missing data mechanisms}\label{fg::DAG}
\end{figure}

In previous papers (\citealp{Frangakis1999}; \citealp{Malley2005}; \citealp{Zhou2006}), the
LI assumption is used for modeling missing data, which means that the potential outcomes and associated potential nonresponse indicators are independent within each principal stratum, that is $P\{ R(z)\mid Y(z),D(z),U\}=P\{ R(z) \mid U\}$. Under the ODN missing data mechanism, the missing data indicator
depends on the possibly missing outcome $Y$, which may be more
reasonable than the LI missing data assumption in some applications.
For example, some patients may have higher probabilities to leave the trial if their health outcomes are not good, and they may be more likely to stay in the trial otherwise. 
We illustrate the LI and ODN missing data mechanisms using the graphical models in Figure \ref{fg::DAG}. Note that the arrows from $Z$ to $R$ are absent because of the compound exclusion restriction assumption. 

\section{Semiparametric Identifiability and Estimation}   \label{sec::semipara}

In this section, we first discuss the difficulty of nonparametric identifiability without assuming a parametric form for both the outcome distribution and the missing data mechanism. If both the distribution of the outcome $Y$
and the missing data mechanism $\rho(y)$ are not specified, the model
is essentially not identifiable without further assumptions. We then
propose a semiparametric method,
specifying only the distribution of $Y$ without assuming any parametric
form for the missing data mechanism. We show the identifiability and
propose a TSMLE of $CACE(Z\rightarrow Y)$ under the assumption that
the distribution of the outcome variable $Y$ belongs to the
exponential family.

\subsection{Semiparametric Identifiability}  \label{subsec::semipara}

Under SUTVA, randomization and monotonicity assumptions, we have
 $\xi=P(Z=1)$, $\omega_a = P(U=a)=P(D=1\mid Z=0)$, $\omega_n = P(U=n)
= P(D=0\mid Z=1)$, and $\omega_c = 1-\omega_a - \omega_n$. These
parameters can be identified directly from the observed data. Next
we focus on the identification of the parameters of $Y$.

\begin{description}
\item[Assumption 7:]
The conditional
density of the outcome variable $Y$ belongs to the following
exponential family:
\begin{equation} \label{eq::expon}
f_{zu}(y) =
f(y\mid Z=z,U=u) = c(\theta_{zu}) h(y) \exp\left\{
\sum_{k=1}^{K} p_k(\theta_{zu}) T_k(y) \right\},
\end{equation}
where $c(\cdot), h(\cdot), p_k(\cdot)$, and $T_k(\cdot)$ are known
functions, and $\theta= \{\theta_{zu} :z=0,1; u=c,a,n\}$
are unknown parameters. We denote $f(y\mid Z=z,U=u)$ simply as
$f_{zu}$ hereinafter.
\end{description}

The parametric assumption of the outcome is untestable in general, since the missing data mechanism depends arbitrarily on the outcome. But for binary outcome, \citet{Small2009} proposed a goodness-of-fit test for the model under the ODN missing data mechanism.
When the randomization assumption holds, the $CACE$ is the difference
between the expectations of the conditional density of $Y$, that is
$$
CACE=E(Y\mid Z=1,U=c)-E(Y\mid Z=0,U=c)=\int y f_{1c} (y)dy-\int y f_{0c} (y) dy.
$$
Hence
if the parameters of $f_{zu} (y)$ are identified, the $CACE$ is also
identified. The exponential family defined by Assumption 7 includes
many common distributions, such as normal distributions $N(\mu_{zu}, \sigma^2)$, exponential distributions $Exp(\lambda_{zu})$ with mean parameter $1/\lambda_{zu}$, Gamma distributions $Gamma(\alpha_{zu}, \lambda)$ with shape parameter $\alpha_{zu}$ and rate parameter $\lambda$, and the log-normal distributions $Lognormal(\mu_{zu}, \sigma^2)$, where
$CACE$s are specified as $CACE_{nor}=\mu_{1c}-\mu_{0c}$,
$CACE_{exp}=1/\lambda_{1c}-1/\lambda_{0c}$, $CACE_{gam}=\alpha_{1c}
/\lambda-\alpha_{0c} /\lambda$, and $CACE_{log}=\exp
\{\mu_{1c}+\sigma^2/2\}-\exp \{\mu_{0c}+\sigma^2/2\}$, respectively.

Next, Theorem \ref{thm::semi} will show
the identifiability of the parameters of $\theta$. The proof of Theorem \ref{thm::semi} is provided in Appendix A.
Assumption 5 implies $\theta_{1n}
= \theta_{0n}$ and $ \theta_{1a} = \theta_{0a}$, which can be simplified as $\theta_n$ and $\theta_a,$ respectively.

\begin{theorem} \label{thm::semi}
Under Assumptions 1 to 7, the vector $\eta =  ( p_1(\theta_n) - p_1(\theta_a)$, $ p_1(\theta_{1c}) -
p_1(\theta_a)$, $p_1(\theta_{0c}) - p_1(\theta_n)$, $\ldots$,
$
 p_K(\theta_n) -  p_K(\theta_a)$,
$ p_K(\theta_{1c}) -  p_K(\theta_a)$, $
p_K(\theta_{0c}) -  p_K(\theta_n)$, $ \log\left\{
c(\theta_n)\right\} - \log\left\{ c(\theta_a)
\right\}$, $\log\left\{ c(\theta_{1c})\right\} - \log\left\{
c(\theta_a)  \right\}$, $ \log\left\{
c(\theta_{0c})\right\} - \log\left\{ c(\theta_n)
\right\}  ) $ is identifiable. If there exists a one-to-one mapping from the parameter set $\theta$ to the vector
$\eta$, then $\theta$ is identifiable and so is $CACE.$
\end{theorem}

 The one-to-one mapping condition seems complicated,
but it is reasonable and holds for many widely-used distributions,
such as homoskedastic normal distributions, exponential
distributions, etc. We will verify the one-to-one mapping condition for
normal and exponential distributions in Appendix C and Appendix D. Other distributions such as heteroskedastic
normal distributions, Gamma distributions and lognormal
distributions can be
verified similarly. However, counterexamples do exist, and we provide one in Appendix A.

\subsection{TSMLE of $CACE$}  \label{subsec::semi}
Because we do not specify a parametric form on the missing data
mechanism $\rho (y) $, the joint distribution of $(Z,U,D,Y,R)$ is not
specified completely. Thus the MLEs of parameters are hard to obtain,
since the likelihood depends on the infinite dimensional parameter $\rho(y)$ as shown in Appendix B.
In this subsection, we propose a two-step likelihood method to
estimate parameters, which can be viewed as an example of the Two-Step
Maximum Likelihood studied by \citet{Murphy2002}.

In the first step, a consistent estimator for $\alpha= (\xi,
\omega_a, \omega_n)$ can be obtained by MLE using the data $\{ (Z_i,
D_i) : i=1,...,N\}$. Let $N$ denote the sample size, $N_1 =
\#\{i:Z_i=1\}$, $N_0 = \#\{i:Z_i=0\}$ and $n_{zd} = \#\{i:Z_i=z, D_i=d \}$
for $z=0,1$ and $d=0,1$. Then the log likelihood function for $\alpha$
is
\begin{equation}\label{eq::like1}
l_1(\alpha) = N_1 \log\xi + N_0\log(1 - \xi) + n_{11}\log(1 -
\omega_n) + n_{10} \log(\omega_n) + n_{01}\log (\omega_a) +
n_{00}\log(1- \omega_a).
\end{equation}
The MLE for $\alpha$ is $\hat{\alpha} = (\hat{\xi}
,\hat{\omega}_n, \hat{\omega}_a ) = (N_1/N , n_{10}/N_1,
n_{01}/N_0)$, equivalent to the moment estimator.

In the second step, we propose a conditional likelihood method to
estimate the parameter set $\theta $, which is based on the
conditional probability of $(Z,D)$ given $Y$ and $R=1$. Here the
proposed conditional likelihood function does not depend on the
nuisance parameter $\rho(y)$, based on the fact that the following
equations (\ref{eq::semi1}) to (\ref{eq::semi3}) do not depend on
$\rho(y)$:
\begin{eqnarray}
& &\log\left\{   \frac{P(Z=1,D=0\mid Y=y,R=1)(1-\xi)}{P(Z=0,D=1\mid Y=y,R=1)\xi}       \right\}   \nonumber\\
&=&\log\left\{   \frac{P(U=n)f(y\mid Z=1,U=n)}{P(U=a)f(y\mid Z=0,U=a)}        \right\} \nonumber\\
&=&\sum_{k=1}^K   \left\{  p_k(\theta_n) -
p_k(\theta_a)  \right\} T_k(y) + \log\left\{ \frac{\omega_n
c(\theta_n)}{\omega_a c(\theta_a)} \right\},
\label{eq::semi1}\\
& &\log\left\{   \frac{P(Z=0,D=0\mid Y=y,R=1)\xi}{P(Z=1,D=0\mid Y=y,R=1)(1-\xi)}    -1   \right\}   \nonumber\\
&=&\log\left\{   \frac{P(U=n)f(y\mid Z=1,U=n) + P(U=c)f(y\mid Z=0,U=c)}{P(U=n)f(y\mid Z=1,U=n)}   -1     \right\}   \nonumber \\
&=&\log\left\{   \frac{ P(U=c)f(y\mid Z=0,U=c)}{P(U=n)f(y\mid Z=1,U=n)}      \right\}   \nonumber \\
&=&\sum_{k=1}^K \left\{ p_k(\theta_{0c}) -
p_k(\theta_n)\right\} T_k(y) + \log\left\{ \frac{\omega_c
c(\theta_{0c})}{\omega_a c(\theta_n)} \right\} ,
\label{eq::semi2}\\
& &\log\left\{   \frac{P(Z=1,D=1\mid Y=y,R=1)(1-\xi)}{P(Z=0,D=1\mid Y=y,R=1)\xi}    -1   \right\}  \nonumber\\
&=&\log\left\{   \frac{P(U=a)f(y\mid Z=1,U=a) + P(U=c)f(y\mid Z=1,U=c)}{P(U=a)f(y\mid Z=0,U=a)}   -1     \right\} \nonumber\\
&=&\log\left\{   \frac{P(U=c)f(y\mid Z=1,U=c)}{P(U=a)f(y\mid Z=0,U=a)}      \right\} \nonumber\\
&=&\sum_{k=1}^K \left\{ p_k(\theta_{1c}) -
p_k(\theta_a) \right\} T_k(y) + \log\left\{ \frac{\omega_c
c(\theta_{1c})}{\omega_a c(\theta_a)} \right\} .
\label{eq::semi3}
\end{eqnarray}
It is obvious that
\begin{equation}\label{eq::semi4}
\sum\limits_{z=0}^1 \sum\limits_{d=0}^1 P(Z=z,D=d\mid Y=y,R=1)=1.
\end{equation}

The left hand sides of equations \eqref{eq::semi1} to
\eqref{eq::semi3} consist of $ P(Z=z,D=d\mid  Y=y,R=1)$ and $\xi$, with the latter identified from the first step. The right hand sides of
equations \eqref{eq::semi1} to \eqref{eq::semi3} consist of the
parameters of interest. Therefore we can estimate $\theta$ through a likelihood
method. Note the right hand sides do not depend on $\rho(y)$, so we do
not need to specify the form of $\rho(y)$. Let $p_{zd}(\theta,
\alpha;y)$  denote $ P(Z=z,D=d\mid Y=y,R=1)$. Since $(Z,D)$
given $(Y=y,R=1)$ follows a multinomial distribution with four
categories, the conditional log-likelihood function of $(Z,D)$ can
be written as
\begin{equation} \label{eq::logL}
l_2(\theta, \alpha) = \sum\limits_{i=1}^N \sum\limits_{z=0}^1 \sum\limits_{d=0}^1 I(Z_i=z,D_i=d, R_i=1)\log p_{zd}(\theta, \alpha;y_i).
\end{equation}

From the proof of Theorem \ref{thm::semi}, the parameter $\theta$ can be
identified from the second likelihood function (\ref{eq::logL})
after identifying $\alpha$ from the first likelihood function
\eqref{eq::like1}. Therefore, by maximizing $l_2(\theta,
\hat{\alpha})$ over $\theta$, we obtain the maximizer $\hat{\theta}$.
In practice, we can use the bootstrap method to approximate the sampling variance of the estimator of $CACE.$

\section{Simulation Studies and Sensitivity Analysis}   \label{sec::simulation}
We report simulation studies and sensitivity analyses in
order to evaluate the finite sample properties of the estimating
methods proposed in this paper.
In Tables \ref{tb::simulation_semip}-\ref{tb::comparison}, the columns
with labels ``bias'', ``$Std.$
$dev.$'', ``$95 \%$ CP'' and ``$95 \%$ CI''
represent the average bias, standard deviation, $95 \%$ coverage
proportion and average $95 \%$ confidence interval, respectively.

First, we generate the outcomes under the ODN missing data
mechanism from homoskedastic normal distributions (denoted as
``$homo\_ normal$''), exponential
distributions, Gamma distributions and log-normal distributions,
respectively. We set the number of simulations to be $10000$, and choose
the sample sizes as $500, 1000$, $2000$ and $4000$, respectively. We show the joint distributions of $(Z, U, D, Y,
R)$ in Table \ref{tb::true_para}, and
report the results in Table \ref{tb::simulation_semip}. The results
have small biases and standard deviations, which decrease as the
sample sizes become larger.
And all the confidence intervals of $CACE$ have empirical coverage proportions very close to their nominal values.

\begin{table}
\begin{center}
\caption{True parameters for simulation}\label{tb::true_para}
\begin{tabular}{l|llll} \hline
    &$homo\_ normal$ & Exponential& Gamma&Lognormal\\ \hline
$Y(1)|U=c$&$N(5,1)$&$Exp(1/5)$&$Gamma(5,1)$&$Lognormal(0,1)$\\
$Y(0)|U=c$&$N(4,1)$&$Exp(1/4)$&$Gamma(4,1)$&$Lognormal(-1,1)$\\
$Y(z)|U=a$&$N(6,1)$&$Exp(1/6)$&$Gamma(6,1)$&$Lognormal(-1.5,1)$\\
$Y(z)|U=n$&$N(3,1)$&$Exp(1/3)$&$Gamma(3,1)$&$Lognormal(-0.5,1)$\\
$Z$&\multicolumn{4}{l}{Bernoulli(0.5) } \\
$U$& \multicolumn{4}{l}{$\omega_c=\omega_n=\omega_a =1/3$ } \\
$R|Y=y$& \multicolumn{4}{l}{$\rho_y = I(y\leq 2) \times 0.85 + I(y\geq7)\times 0.8 + I(2<y<7)\times 0.9$}\\ \hline
\end{tabular}
\end{center}
\end{table}

\begin{table}
\begin{center}
\caption{Results of simulation studies %without covariates: SemiP
}\label{tb::simulation_semip}
\begin{tabular}{llrrrr} \hline
    \multicolumn{1}{c}{true value}    & \multicolumn{1}{c}{$N$ }& \multicolumn{1}{c}{bias } & \multicolumn{1}{c}{$Std.$ $dev.$ } & \multicolumn{1}{c}{$ 95 \%$ CP} & \multicolumn{1}{c}{$ 95 \%$ CI} \\ \hline
         $CACE_{nor}=1.0$& 500   &   -0.0194  &   0.3395  & 0.9489 &[0.3152, 1.6461]\\
          & 1000  &   -0.0073 &   0.2343  & 0.9476& [0.5335, 1.4518]\\
                    & 2000  &   0.0022 &   0.1629  & 0.9500 & [0.6828, 1.3215]\\
                    & 4000  &   -0.0019&   0.1145  & 0.9504 & [0.7736, 1.2225]\\ \hline
    $CACE_{exp}=1.0$& 500   &   0.0872   &   1.5910 & 0.9455 & [-2.0312, 4.2056]\\
          & 1000  &   0.0312   &   1.0309 & 0.9441&[-0.9893, 3.0517]\\
                    & 2000  &   0.0106   &   0.7091 & 0.9479&[-0.3793, 2.4004] \\
                    & 4000  &   0.0072   &   0.4891 & 0.9506&[0.0486, 1.9657] \\ \hline
         $CACE_{gam}=1.0$&  500   &   0.0830    &   1.6872 & 0.9915&[-2.2237, 4.3901]\\
          & 1000  &   0.0284   &   0.5978 & 0.9625&[-0.1432, 2.2000]\\
                    & 2000  &   0.0108   &   0.3636 & 0.9493&[0.2981, 1.7235] \\
                    & 4000  &   0.0032   &   0.2530 & 0.9505 &[0.5073, 1.4992]\\ \hline
      $CACE_{log}=1.0422$& 500   &  0.1156   & 0.7849    & 0.9617 &[-0.3806, 2.6962]\\
         & 1000  &  0.0599  &  0.4571   & 0.9494 &[0.2061, 1.9981]\\
                    & 2000  & 0.0218  &  0.3093   &  0.9496&[0.4578, 1.6702]\\
                    & 4000  & 0.0106  &  0.2130  & 0.9469 &[0.6353, 1.4702] \\ \hline
  %       \multicolumn{5}{c}{ Results of ``with covariates'' }\\ \hline
  %        SemiP& bias  &  -0.023 &   -0.017 &  -0.003\\
  %                  & mse  &   0.331 &   0.051  &  0.023\\
  %                  & cp     &   0.944 &    0.949 &  0.963\\ \hline
  %        SemiPX&bias &   0.027 &   0.011 &   0.012\\
  %                  &mse   &   0.129 &   0.042  &  0.018\\
  %                  &cp      &   0.971 &   0.982 &    0.988 \\ \hline \hline
\end{tabular}
\end{center}
\end{table}

Second, we report the results of comparison of our methods with the MLE proposed by O'Malley and Normand (2005)
 (``LI'' in Tables \ref{tb::sensi}) 
under five
different cases, which violate the homoskedastic normal outcomes or the ODN assumption. We repeat our simulation $10000$ times
with sample sizes of $4000$ in each case. The results of five
cases are shown in Table \ref{tb::sensi}, named as ``Heter'', 
``Unif'',``T'', ``DY'' and ``DYU'', respectively. The first
case, ``Heter'' case, violates the homoskedastic normal outcomes. Next two cases, ``Unif'' and ``T'', violate
the exponential family assumption. The last two cases, ``DY'' and
``DYU'', violate the ODN assumption. In the ``Heter'' case we
generate data from heteroskadastic normal outcomes. The data
generating process is the same as ``$homo\_ normal$'' except
that $Y(1)|U=c\sim N(5, 0.25)$, $Y(z)|U=a \sim N(6, 0.30)$ for
$z=0,1$. In the `Unif'' case the data is generated the same as
``$homo\_ normal$'' except that the outcomes follow uniform
distribution with $Y(1)|U=c \sim U[2,8]$, $Y(0)|U=c \sim U[1,7]$,
$Y(z)|U=a \sim U[3,9]$ and $Y(z)|U=n \sim U[1,5]$, respectively. 
The data generating process in the ``T'' case is the same as
``$homo\_ normal$'' except that the outcomes follow t
distributions with the same means as ``$homo\_ normal$'' and
degrees of freedom $4$. In the ``DY'' case we generate data under
the missing data mechanism depending on both $D$ and $Y$, and choose
$P(R=1\mid D, Y) = 0.8 - I(Y > 5)\times 0.5 + I(D=1)\times 0.1
- I(Y
> 5)I(D=0)\times 0.1$ with other conditional distributions the same
as ``$homo\_ normal$''. In the ``DYU'' case we generate data
with the missing data mechanism depending on $D$, $Y$ and $U$, and
choose $P(R=1\mid D, Y, U) = (1 + \exp\{5+  0.1D - Y -
0.1U\})^{-1}$, with other conditional distributions the same as
``$homo\_ normal$'' and $U=1, 2, 3$ corresponding to $U=c,n,a$.
From Table \ref{tb::sensi}, we can see that the point estimator of our method is generally
robust to four kinds of violations of the assumptions.
However, the results are worse for ``Unif'' case, which has a large
bias, low $95\%$ coverage proportion and whose $95\%$ confidence
interval does not cover the true value.

\begin{table}
\begin{center}
\caption{Comparison of the methods assuming ODN and LI under five
cases which violate the homoskedastic normal outcomes or the ODN assumption.
($CACE_{true}=1.0$) }\label{tb::sensi}
\begin{tabular}{c|l|rrrr}\hline
  \multicolumn{1}{c|}{Method}  &  Assump.    & \multicolumn{1}{c}{bias}& \multicolumn{1}{c}{$Std.$ $dev.$ } & \multicolumn{1}{c}{$ 95 \%$ CP}  & \multicolumn{1}{c}{$ 95 \%$ CI}  \\ \hline
 ODN &  Heter &  -0.0268  & 0.0772  &0.9363   & [0.8220, 1.1244] \\
    &Unif&-0.3815&0.1865&0.4676&[0.2530, 0.9841] \\
    &T        &  -0.0350   & 0.1730 & 0.9427 & [0.6960, 1.3740]\\
    &DY &0.0201&0.1555&0.9465&[0.7154, 1.3249] \\
    & DYU &-0.0852&0.1691&0.9242&[0.5834, 1.2462]\\ \hline
 LI & Heter& -0.0277&0.0677&0.9300&[0.8395, 1.1051] \\
    & Unif &-0.8521 & 0.2474 & 0.0695 & [-0.3369, 0.6327] \\
    & T & 0.2244 & 0.1225 & 0.5577 & [0.9843, 1.4646] \\
    & DY & -0.0288 & 0.0894 & 0.9370 & [0.7959, 1.1465] \\
    & DYU & -0.1267& 0.1321 & 0.8426 & [0.6144, 1.1322] \\ \hline
\end{tabular}
\end{center}
\end{table}

Finally, we compare our methods with the MLE proposed by O'Malley and Normand (2005)
 under the LI missing data mechanism (``LI'' in
Table \ref{tb::comparison}). We repeat our simulation $10000$ times with
sample sizes of $4000$ in each case. The data generating processes are the
same as ``$homo\_ normal$'', but the missing data mechanisms are
LI. Denote $\gamma_{du} = P(R=1\mid D=d, U=u)$, and choose
$(\gamma_{1c}, \gamma_{0c}, \gamma_{0n}, \gamma_{1a}) = (0.8, 0.75,
0.7, 0.9)$, $(0.9,  0.7,  0.8, 0.7)$, $(0.7,0.6,  0.6, 0.8)$ and
$(0.6,0.7,0.9,0.7)$ for ``LI1'' to ``LI4'' respectively as shown
in rows 1-4 and 5-8 of Table \ref{tb::comparison}.
Since the missing data mechanisms are LI, the ``LI'' method exhibits very
small biases. Although the assumptions required by the ``ODN'' methods
do not hold, the biases are not very large except for the missing mechanism LI4.  The last case, LI4, has the largest variability among the $\gamma_{zu}$'s and thus the largest bias for estimating the $CACE$, since the missing data mechanism has the ``strongest'' dependence on $D$ and $U$ but not $Y$. Next we generate data under the ODN
assumption, and compare the methods under both ODN and LI assumptions.
Let $\rho(y;\delta) = I(y\leq 2)
\times (0.9-\delta ) + I(y\geq 7)\times (0.9-2\delta ) +
I(2<y<7)\times 0.9$ where $0<\delta<0.9$. As $\delta$ increases, the relationship between $Y$ and $R$ becomes stronger.
The data are generated from the same joint distribution as ``$homo\_ normal$'' except for different $\rho(y;\delta)$.
The results are shown in Figure 2. The method under ODN missing data mechanism has small bias and
promising coverage property irrespective of $\delta$, but the method under LI missing data mechanism has larger bias and poorer coverage property with larger $\delta$. 

\begin{table}
\begin{center}
\caption{Comparison of the methods assuming ODN and LI when LI holds. ($CACE_{true}=1.0$}\label{tb::comparison}
\begin{tabular}{c|l|rrrr}\hline
Method    &   & \multicolumn{1}{c}{bias}& \multicolumn{1}{c}{$Std.$ $dev.$ } & \multicolumn{1}{c}{$ 95 \%$ CP}  & \multicolumn{1}{c}{$ 95 \%$ CI}\\ \hline
ODN&  LI1&-0.0291&0.1236&0.9438&[0.7287, 1.2132]\\
  &LI2& 0.0976&0.1452&0.8961&[0.8130, 1.3823] \\
  &LI3& -0.0569&0.1371&0.9332&[0.6745, 1.2118]\\
  &LI4&-0.1961&0.1152&0.5966&[0.5781, 1.0297]\\ \hline
LI&LI1&-0.0013&0.1123&0.9491&[0.7785, 1.2189]\\
  &LI2&-0.0010&0.1099&0.9504&[0.7836, 1.2145] \\
  &LI3&-0.0008&0.1262&0.9502&[0.7519, 1.2465]\\
  &LI4&-0.0015&0.1290&0.9505&[0.7456, 1.2515] \\ \hline
\end{tabular}
\end{center}
\end{table}

\begin{figure}
\includegraphics[width = \textwidth]{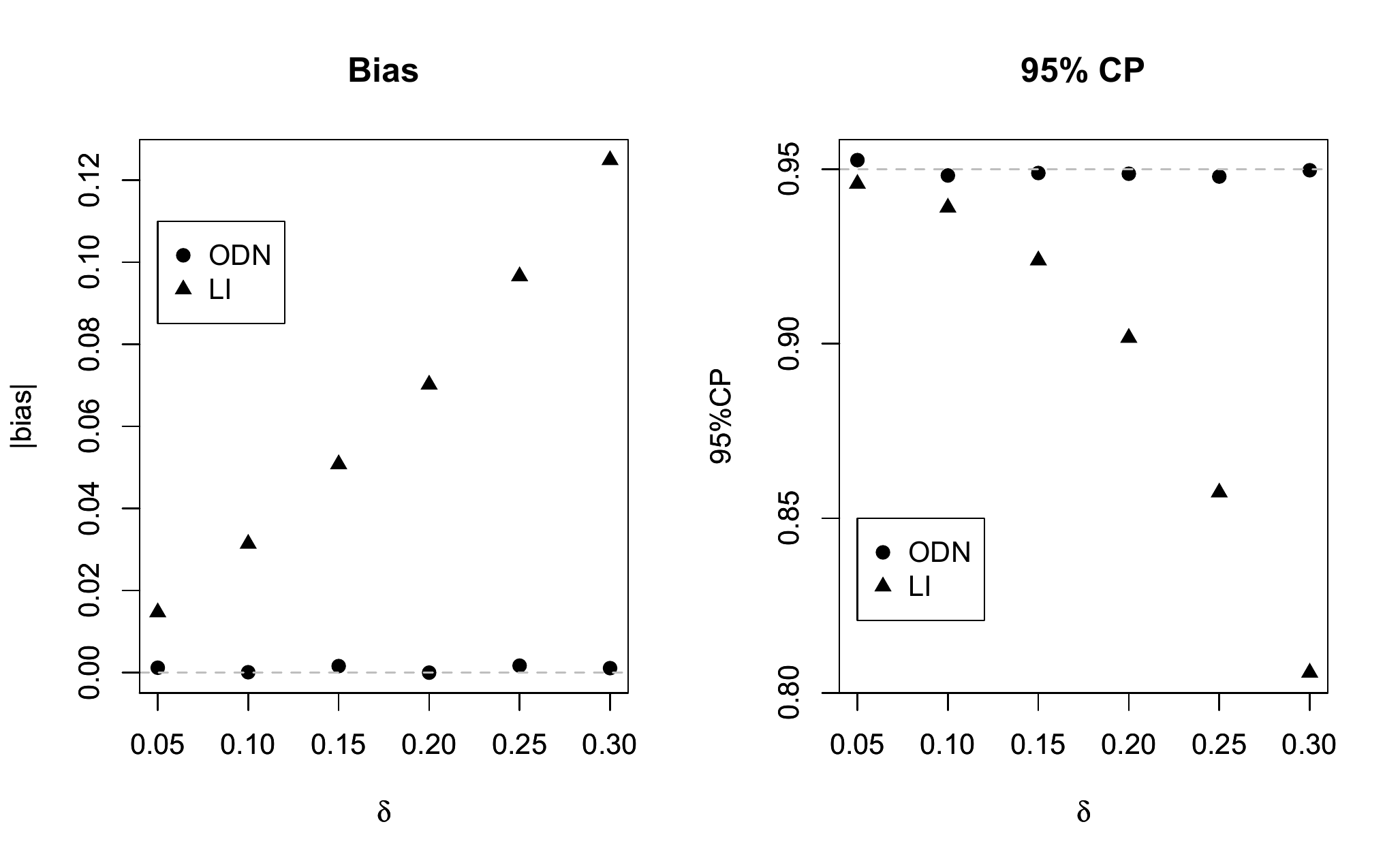}
\caption{Comparison of the methods assuming ODN and LI when ODN holds. ($CACE_{true}=1.0$)}\label{fg::table5}
\end{figure}

\section{Application}  \label{sec::data}
We use the new methods proposed in this paper to re-analyze a
psychiatric clinical trial. It is a double-blinded randomized study
comparing the relative effect of clozapine and haloperidol in adults
with refractory schizophrenia at fifteen Veterans Affairs medical
centers. Clozapine is found to be more efficacious than standard
drugs  in patients with refractory schizophrenia. Yet it is
associated with potentially fatal agranulocytosis. One objective for
conducting this trial is to evaluate the clinical effect of two
antipsychotic medications. The dataset has been analysed in \citet{Rosenheck1997}, \citet{Levy2004}, and
\citet{Malley2005}. Some summary statistics of the data are described in Table \ref{tb::datasummary}.
More details
about the trial can be found in \citet{Rosenheck1997} and
\citet{Malley2005}.
In the treatment arm, $203$ patients
are randomized to clozapine; in the control arm, $218$ patients
are randomized to haloperidol. The
outcome of interest is the positive and negative syndrome score
(PANSS) with higher values indicating more severe symptoms. The baseline PANSS is nearly balanced in both groups.
Missing outcome patterns are obviously different in the clozapine group
(about $40 / 203 \approx 0.20$) than in the haloperidol group
(about $ 59 / 218 \approx 0.27$). Hence it is possible that
outcomes are not missing at random. The primary reasons for
dropout in the clozapine group are side effects or
non-drug-related reasons. The reasons for discontinuing haloperidol
are lack of efficacy or worsening of symptoms. Therefore, the
missing mechanism may possibly depend on the missing outcome, and we think that the
ODN assumption may be more reasonable in this case.

\begin{table}[ht]
\begin{center}
\caption{Summary statistics of the data from the psychiatric clinical trial}\label{tb::datasummary}
\begin{tabular}{lccc}\hline
& \multicolumn{2}{c}{Received treatment } &\\ \cline{2-3}
Assigned treatment        &Clozapine ($D=1$)&Haloperidol ($D=0$)&Total \\ \hline
  Clozapine ($Z=1$)      & &&\\
 \hspace{4mm}Sample size       &  122&81&203 \\
  \hspace{4mm}Missing sample size& 0  &40&40\\
  \hspace{4mm}Mean of the baseline PANSS&   90.83&91.20&90.98\\
  Haloperidol ($Z=0$)     & &&\\
 \hspace{4mm}Sample size      & 57&161 & 218 \\
  \hspace{4mm}Missing sample size&  12 &47&59\\
  \hspace{4mm}Mean of the Baseline PANSS &96.30&90.69&   92.16 \\ \hline
\end{tabular}
\end{center}
\end{table}

The estimates of $CACE$ by different methods are shown in Table \ref{tb::data}. In Table \ref{tb::data},
the ``homo'' and ``hetero'' in parentheses after ``ODN''
correspond to the homoskedastic and heteroskedastic model
assumptions, respectively; and ``LI'' corresponds to the MLE proposed in
\citet{Malley2005}.
The columns of Table \ref{tb::data} correspond to the methods, point estimates, standard errors, $95\%$ and $90\%$ confidence intervals, respectively. The bootstrap method is used to compute standard errors and confidence intervals for all methods.
From Table \ref{tb::data}, we can see
that under the homoskedastic assumption subjects in the clozapine group
had $5.00$ lower symptom levels than those in the haloperidol group,
and under the heteroskedastic assumption it was $5.54$. Both methods have
similar conclusions that clozapine is somewhat more effective than
haloperidol for patients with refractory schizophrenia. Both of
the semiparametric methods give insignificant results since both
$95\%$ confidence intervals include zero. Our results are similar
to \citet{Malley2005} which also gave an insignificant result. However, the results of the $90\%$ confidence intervals are somewhat different: the result under the ODN mechanism with the heteroskedastic assumption is significant, but the results from other two models are not.

Assuming different missing data mechanisms such as LI, ODN and under
other different assumptions on the outcome variable, we can find
different point estimates and confidence intervals for $CACE$ using the data from the psychiatric clinical trial.
When we have prior knowledge that the missing mechanism depends only on the treatment received and the compliance status, the method under LI missing mechanism will provide more credible conclusion.
However, when we have prior knowledge that the missing mechanism depends directly on the outcome, we recommend our methods under the ODN missing data mechanism.
The newly proposed methods can be used as alternatives for the predominate methods assuming the LI missing mechanism in sensitivity analysis.

\begin{table}[ht]
\begin{center}
\caption{Estimates of $CACE$ by different methods}\label{tb::data}
\begin{tabular}{lcccr}\hline
    Method     & Estimate & $S.E.$ & $95\%$ CI &\multicolumn{1}{c}{ $90\%$ CI}
    \\ \hline
    ODN(homo)&      -5.00 & 3.05 & [-10.98, 0.98]&[-10.02, 0.02]\\
    ODN(hetero)&      -5.54 & 3.05 & [-11.52, 0.44]&[-10.56, -0.52] \\
    LI &-4.36 & 4.35& [-12.89, 4.17]&[-11.52, 2.80]\\ \hline
\end{tabular}
\end{center}
\end{table}

\section{Discussion}   \label{sec::discuss}
Randomization is a powerful tool to measure the relative causal effect
of treatment versus control. Some subjects in randomized
trials, however, may fail to comply with the assigned treatments or drop out before the final outcomes are measured.
Noncompliance and missing data problems make statistical causal
inference difficult, because the causal effects are not identifiable
without additional assumptions. Under different assumptions about the missing data mechanisms of the outcomes, the
identifiability and estimation methods may be fundamentally
different. Most previous studies (\citealp{Frangakis1999}; \citealp{Barnard2003}; \citealp{Malley2005}; \citealp{Zhou2006}) rely on the LI assumption in order to identify $CACE$, but
the LI assumption may be not reasonable when the missing data
mechanism may depend on the outcome. Under the ODN missing data
mechanism, \citet{Chen2009} and \citet{Imai2009} showed the identifiability
and proposed the moment estimator and the MLE of $CACE$ for discrete outcomes. But
there are no results for continuous outcomes under both
noncompliance and ODN missing data mechanism. As a generalization of
\citet{Chen2009} and \citet{Imai2009}, we study the
semiparametric identifiability, and propose estimation methods for
$CACE$ with continuous outcomes under the ODN missing data mechanism.

The ODN assumption allows the missing data mechanism to
depend on the outcome. However, the missing data processes in
practical problems may be more complicated, and they may depend on
other variables such as $Z$, $U$ and $D$. For example, a missing mechanism depending on both the compliance status and the outcome may be reasonable in some real studies.
\citet{Small2009} proposed a saturated model for $P(R=1\mid Z,U,Y)$, and the
models under LI and ODN are special cases of their model. However,
their model is generally not identifiable without restrictions on the
parameters. It is worthwhile to study the identifiability of $CACE$
under all possible restrictions of $P(R=1\mid Z,U,Y)$ and perform
sensitivity analysis for models lack of identifiability. We
consider only cross-sectional data in this paper, and generalizing our
methods to longitudinal data is a future research topic.

\section*{Acknowledgments}
We would like to thank Editor, Associate Editor and three reviewers
for their very valuable comments and suggestions.
Chen's research was supported in part by 
NSFC 11101045, CAEP 2012A0201011 and CAEP 2013A0101004. Geng's
research was supported by NSFC 11021463, 10931002 and 11171365.
Zhou's research was supported in part by Department of Veterans
Affairs HSR\&D RCS Award 05-196. It does not necessarily represent
the views of VA HSR\&D Service.

\section*{Appendices}

\subsection*{Appendix A}
%set the counter of equations
\setcounter{equation}{0}
\renewcommand {\theequation} {A.\arabic{equation}}

\noindent {\it Proof of Theorem 3.1.}
From equations (3.5) to (3.7), we can
identify $( p_1(\theta_n) -  p_1(\theta_a)$, $
p_1(\theta_{1c}) -  p_1(\theta_a)$, $
p_1(\theta_{0c}) -  p_1(\theta_n)$, $ \ldots$, $
p_K(\theta_n) -  p_K(\theta_a)$, $
p_K(\theta_{1c}) -  p_K(\theta_a)$, $
p_K(\theta_{0c}) -  p_K(\theta_n)$, $ \log\left\{
c(\theta_n)\right\} - \log\left\{ c(\theta_a)
\right\} $, $ \log\left\{ c(\theta_{1c})\right\} - \log\left\{
c(\theta_a)  \right\}$,
 $ \log\left\{ c(\theta_{0c})\right\} - \log\left\{ c(\theta_n)  \right\}  )$
using generalized linear models.
Therefore
$\theta$
is identifiable because of the one-to-one mapping.

\noindent {\it Counterexample for identifiability.}
Consider the following exponential family: 
$$
f_{zu}(y) = f(y\mid Z=z, U=u) = c(\theta_{zu}) h(y) 
\exp\left\{   \sum_{k=1}^K \theta_{zu,k} y^k      \right\}.
$$
The number of unknown parameters contained in $\theta$ is $4K$, and the number of identifiable parameters contained in $\eta$ is $3(K+1)$. A necessary condition for the existence of a one-to-one mapping from $\theta$ to $\eta$ is $4K\leq 3(K+1)$, or, equivalently, $K\leq 3.$ Therefore, when $K>3$, a one-to-one mapping from $\theta$ to $\eta$ does not exist.

\subsection*{Appendix B: Full likelihood for $(\alpha, \theta, \rho(y))$}
Define $M_{zd} = \#\{ i: Z_i=z, D_i =d, R_i=0  \}$ for $z=0,1$ and $d=0,1.$ Under the compound exclusion restriction, we have $f_{1n} (y) = f_{0n} (y) = f_n(y)$ and $f_{1a}(y) = f_{0a}(y) = f_a(y).$
The full likelihood for $(\alpha, \theta, \rho(y))$ is
\begin{eqnarray*}
&&L(\alpha, \theta, \rho(y))\\
&\propto& \xi^{N_1} (1 - \xi)^{N_0}  (1 - \omega_n)^{  n_{11} }  \omega_n^{n_{10}}  \omega_a^{n_{01}}  (1- \omega_a)^{n_{00}} \\
&&\cdot     \left[   \frac{  \omega_c  }{  \omega_c + \omega_a } \int \{ 1 - \rho(y)\} f_{1c}(y)dy +  \frac{  \omega_a  }{  \omega_c + \omega_a } \int \{ 1 - \rho(y)\} f_{a}(y)dy \right] ^{ M_{11} }  \left[  \int  \{ 1- \rho(y) \} f_{n}(y)dy \right]^{ M_{10} }\\
&&\cdot \left[   \int  \{ 1-\rho(y)\} f_{a}(y)dy \right]^{ M_{01} }  \left[    \frac{  \omega_c  }{  \omega_c + \omega_n } \int \{ 1-\rho(y)\} f_{0c}(y)dy +  \frac{  \omega_n  }{  \omega_c + \omega_n } \int \{ 1-\rho(y)\} f_{n}(y)dy \right]^{ M_{00} }\\
&&\cdot \prod_{i: R_i=1}   \rho(Y_i) \cdot \prod_{i:  (Z_i, D_i, R_i) = (1,1,1)}  \left\{    \frac{  \omega_c  }{  \omega_c + \omega_a }  f_{1c}(Y_i) 
+     \frac{  \omega_a  }{  \omega_c + \omega_a }  f_a(Y_i)  \right\}  
 \prod_{i: (Z_i, D_i, R_i) = (1,0,1)}   f_{n}(Y_i)  \\
&& \cdot \prod_{i: (Z_i, D_i, R_i) = (0,1,1)}  f_a(Y_i) \cdot 
\prod_{i: (Z_i, D_i, R_i) = (0,0,1)}  \left\{   \frac{ \omega_c  }{  \omega_c + \omega_n } f_{0c}(Y_i) + \frac{  \omega_n  }{  \omega_c + \omega_n } f_n(Y_i) \right\}.
\end{eqnarray*}

\subsection*{Appendix C: Verification of homoskedastic normal distribution in Subsection 3.2}

For homoskedastic normal outcomes, equations (3.5) to (3.7) can be re-written as:
\begin{eqnarray*}
a_1 y + b_1& =&\log\left\{   \frac{P(Z=1,D=0\mid Y=y,R=1)(1-\xi)}{P(Z=0,D=1\mid Y=y,R=1)\xi}       \right\}  \\
&=&\frac{\mu_n - \mu_a }{ \sigma^2}y + \frac{\mu_a^2 - \mu_n^2}{2\sigma^2}+\log\left\{ \frac{\omega_n}{\omega_a} \right\},\\
a_2 y + b_2& =&\log\left\{   \frac{P(Z=0,D=0\mid Y=y,R=1)\xi}{P(Z=1,D=0\mid Y=y,R=1)(1-\xi)}    -1   \right\}\\
&=&\frac{\mu_{0c} - \mu_n }{ \sigma^2}y + \frac{\mu_n^2 - \mu_{0c}^2}{2\sigma^2}+\log\left\{ \frac{\omega_c}{\omega_n} \right\},\\
a_3 y + b_3&   =&\log\left\{   \frac{P(Z=1,D=1\mid Y=y,R=1)(1-\xi)}{P(Z=0,D=1\mid Y=y,R=1)\xi}    -1   \right\}   \nonumber\\
&=&\frac{\mu_{1c} - \mu_a }{ \sigma^2}y + \frac{\mu_a^2 - \mu_{1c}^2}{2\sigma^2}+\log\left\{ \frac{\omega_c}{\omega_a} \right\}.\\
\end{eqnarray*}
Since $a_i$ and $b_i$ can be identified from generalized linear models, we can identify all the parameters from the above equations and obtain the following results:
\begin{eqnarray*}
\sigma^2 &=& 2 \left[   \frac{b_1 -\log  \left\{ \frac{\omega_n}{\omega_a} \right\} }{a_1}   -    \frac{b_2 -\log \left\{ \frac{\omega_c}{\omega_n} \right\} }{a_2}               \right]/(a_1 + a_2),\\
&=& 2 \left[   \frac{b_1 -\log  \left\{ \frac{\omega_n}{\omega_a} \right\}}{a_1}   -    \frac{b_3 -\log \left\{ \frac{\omega_c}{\omega_a} \right\}}{a_3}               \right]/(a_3 - a_1),\\
\mu_{1c} &=& \frac{1}{2}a_3 \sigma^2 - \frac{b_3 - \log \left\{ \frac{\omega_c}{\omega_a} \right\}}{a_3},\\
\mu_{0c} &=& \frac{1}{2}a_2 \sigma^2 - \frac{b_2 - \log \left\{ \frac{\omega_c}{\omega_n} \right\}}{a_2}.
\end{eqnarray*}
Therefore, we can identify $ CACE = \mu_{1c} - \mu_{0c}.$

\subsection*{Appendix D: Verification of the exponential distribution in Subsection 3.2}

For the exponentially distributed outcomes, equations
(3.5) to (3.7) can be re-written as:
\begin{eqnarray*}
a_1 y + b_1& =&\log\left\{   \frac{P(Z=1,D=0\mid Y=y,R=1)(1-\xi)}{P(Z=0,D=1\mid Y=y,R=1)\xi}       \right\}  \\
&=&(\lambda_a - \lambda_n)y + \log\left\{  \frac{\omega_n
\lambda_n}{\omega_a \lambda_a} \right\},
\end{eqnarray*}
\begin{eqnarray*}
a_2 y + b_2& =&\log\left\{   \frac{P(Z=0,D=0\mid Y=y,R=1)\xi}{P(Z=1,D=0\mid Y=y,R=1)(1-\xi)}    -1   \right\}   \\
&=&(\lambda_n - \lambda_{0c}) y +\log\left\{
\frac{\omega_c\lambda_{0c}}{\omega_n\lambda_n}  \right\},
\end{eqnarray*}
\begin{eqnarray*}
a_3 y + b_3&  =&\log\left\{   \frac{P(Z=1,D=1\mid Y=y,R=1)(1-\xi)}{P(Z=0,D=1\mid Y=y,R=1)\xi}    -1   \right\} \nonumber\\
&=&(\lambda_a - \lambda_{1c})y + \log\left\{  \frac{\omega_c
\lambda_{1c} }{\omega_a \lambda_a}  \right\}.
\end{eqnarray*}
Since $a_i$ and $b_i$ can be identified from generalized linear models, we can identify all the parameters from the above equations and obtain the following results:
\begin{eqnarray*}
\lambda_{0c}&=&\frac{a_1 \exp \{b_1 - \log\{ \omega_n/\omega_a \}  \} }{ 1-\exp\{ b_1 - \log\{  \omega_n/\omega_a \}  \}  }-a_2 , \\
\lambda_{1c}&=&\frac{a_1  }{ 1-\exp\{ b_1 - \log\{  \omega_n/\omega_a \}  \}  }-a_3 , \\
\text{and }CACE&=& \frac{1}{\lambda_{1c}} - \frac{1}{\lambda_{0c}}.
\end{eqnarray*}

\bibliographystyle{plainnat}
\bibliography{refschen}

\begin{thebibliography}{20}
\providecommand{\natexlab}[1]{#1}
\providecommand{\url}[1]{\texttt{#1}}
\expandafter\ifx\csname urlstyle\endcsname\relax
  \providecommand{\doi}[1]{doi: #1}\else
  \providecommand{\doi}{doi: \begingroup \urlstyle{rm}\Url}\fi

\bibitem[Angrist et~al.(1996)Angrist, Imbens, and Rubin]{Angrist1996}
J.~D. Angrist, G.~W. Imbens, and D.~B. Rubin.
\newblock Identification of causal effects using instrumental variables (with
  discussion).
\newblock \emph{Journal of the American Statistical Association}, 91:\penalty0
  444--455, 1996.

\bibitem[Balke and Pearl(1997)]{Balke1997}
A.~Balke and J.~Pearl.
\newblock Bounds on treatment effects from studies with imperfect compliance.
\newblock \emph{Journal of the American Statistical Association}, 92:\penalty0
  1171--1176, 1997.

\bibitem[Barnard et~al.(2003)Barnard, Frangakis, Hill, and Rubin]{Barnard2003}
J.~Barnard, C.~E. Frangakis, J.~L. Hill, and D.~B. Rubin.
\newblock Principle stratification approach to broken randomized experiments:
  {A} case study of school choice vouchers in {N}ew {Y}ork {C}ity (with
  discussion).
\newblock \emph{Journal of the American Statistical Association}, 98:\penalty0
  299--314, 2003.

\bibitem[Chen et~al.(2009)Chen, Geng, and Zhou]{Chen2009}
H.~Chen, Z.~Geng, and X.~H. Zhou.
\newblock Identifiability and estimation of causal effects in randomized trials
  with noncompliance and completely non-ignorable missing data (with
  discussion).
\newblock \emph{Biometrics}, 65:\penalty0 675--682, 2009.

\bibitem[Dempster et~al.(1977)Dempster, Laird, and Rubin]{Dempster1977}
A.P. Dempster, N.M. Laird, and D.B. Rubin.
\newblock Maximum likelihood from incomplete data via the {EM} algorithm (with
  discussion).
\newblock \emph{Journal of the Royal Statistical Society: Series B (Statistical
  Methodology)}, 39:\penalty0 1--38, 1977.

\bibitem[Efron and Feldman(1991)]{Efron1991}
B.~Efron and D.~Feldman.
\newblock Compliance as an explanatory variable in clinical trials (with
  discussion).
\newblock \emph{Journal of the American Statistical Association}, 86:\penalty0
  9--17, 1991.

\bibitem[Frangakis and Rubin(1999)]{Frangakis1999}
C.~E. Frangakis and D.~B. Rubin.
\newblock Addressing complications of intention-to-treat analysis in the
  combined presence of all-or-none treatment-noncompliance and subsequent
  missing outcomes.
\newblock \emph{Biometrika}, 86:\penalty0 365--379, 1999.

\bibitem[Frangakis and Rubin(2002)]{Frangakis2002}
C.~E. Frangakis and D.~B. Rubin.
\newblock Principal stratification in causal inference.
\newblock \emph{Biometrics}, 58:\penalty0 21--29, 2002.

\bibitem[Hudgens and Halloran(2008)]{Hudgens2008}
M.~G. Hudgens and M.~E. Halloran.
\newblock Toward causal inference with interference.
\newblock \emph{Journal of the American Statistical Association}, 103:\penalty0
  832--842, 2008.

\bibitem[Imai(2009)]{Imai2009}
K.~Imai.
\newblock Statistical analysis of randomized experiments with non-ignorable
  missing binary outcomes: an application to a voting experiment.
\newblock \emph{Journal of the Royal Statistical Society: Series C (Applied
  Statistics)}, 58:\penalty0 83--104, 2009.

\bibitem[Imbens and Rubin(1997)]{Imbens1997}
G.~W. Imbens and D.~B. Rubin.
\newblock Bayesian inference for causal effects in randomized experiments with
  noncompliance.
\newblock \emph{The Annals of Statistics}, 25:\penalty0 305--327, 1997.

\bibitem[Levy et~al.(2004)Levy, O'Malley, and T.]{Levy2004}
D.~E. Levy, A.~J. O'Malley, and Normand S.~L. T.
\newblock Covariate adjustment in clinical trials with non-ignorable missing
  data and non-compliance.
\newblock \emph{Statistics in Medicine}, 23:\penalty0 2319--2339, 2004.

\bibitem[Murphy and Topel(2002)]{Murphy2002}
K.M. Murphy and R.H. Topel.
\newblock Estimation and inference in two-step econometric models.
\newblock \emph{Journal of Business and Economic Statistics}, 20:\penalty0
  88--97, 2002.

\bibitem[O'Malley and Normand(2005)]{Malley2005}
A.~J. O'Malley and S.~L.~T. Normand.
\newblock Likelihood methods for treatment noncompliance and subsequent
  nonresponse in randomized trials.
\newblock \emph{Biometrics}, 61:\penalty0 325--334, 2005.

\bibitem[Rosenheck et~al.(1997)Rosenheck, Cramer, Xu, Thomas, Henderson,
  Frisman, Fye, and Charney]{Rosenheck1997}
R.~Rosenheck, J.~Cramer, W.~Xu, J.~Thomas, W.~Henderson, L.~Frisman, C.~Fye,
  and D.~Charney.
\newblock A comparison of clozapine and haloperidol in hospitalized patients
  with refractory schizophrenia.
\newblock \emph{New England Journal of Medicine}, 337:\penalty0 809--815, 1997.

\bibitem[Rubin(1980)]{Rubin1980}
D.~B. Rubin.
\newblock Comment on ``{R}andomization analysis of experimental data: the
  {F}isher randomization test'' by {D}. {B}asu.
\newblock \emph{Journal of the American Statistical Association}, 75:\penalty0
  591--593, 1980.

\bibitem[Rubin(1986)]{Rubin1986}
D.~B. Rubin.
\newblock Comments on ``{S}tatistics and causal inference'' by {P}aul
  {H}olland: {W}hich ifs have causal answers.
\newblock \emph{Journal of the American Statistical Association}, 81:\penalty0
  961--962, 1986.

\bibitem[Small and Cheng(2009)]{Small2009}
D.~S. Small and J.~Cheng.
\newblock Discussion of ``{I}dentifiability and estimation of causal effects in
  randomized trials with noncompliance and completely non-ignorable missing
  data'' by {H}ua {C}hen, {Z}hi {G}eng and {X}iaohua {Z}hou.
\newblock \emph{Biometrics}, 65:\penalty0 682--685, 2009.

\bibitem[Taylor and Zhou(2011)]{Taylor2011}
L.~Taylor and X.~H. Zhou.
\newblock Methods for clustered encouragement design studies with noncompliance
  and missing data.
\newblock \emph{Biostatistics}, 12:\penalty0 313--326, 2011.

\bibitem[Zhou and Li(2006)]{Zhou2006}
X.~H. Zhou and S.~M. Li.
\newblock I{TT} analysis of randomized encouragement design studies with
  missing data.
\newblock \emph{Statistics in Medicine}, 25:\penalty0 2737--2761, 2006.

\end{thebibliography}

\end{document}